\documentclass[11pt,aps,preprint]{revtex4}
\usepackage[active]{srcltx}
\usepackage[utf8]{inputenc}
\usepackage{latexsym}
\usepackage{amsmath}
\usepackage{graphicx}

\begin{document}

\title{Competing interactions and the Lifshitz-type Nonlinear Sigma Model}

\author{Pedro R. S. Gomes}
\email{pedrorsg@fma.if.usp.br}
\affiliation{Instituto de F\'\i sica, Universidade de S\~ao Paulo\\
Caixa Postal 66318, 05314-970, S\~ao Paulo, SP, Brazil}%

\author{P. F. Bienzobaz}
\email{paulafb@if.usp.br}
\affiliation{Instituto de F\'\i sica, Universidade de S\~ao Paulo\\
Caixa Postal 66318, 05314-970, S\~ao Paulo, SP, Brazil}%

\author{M. Gomes}
\email{mgomes@fma.if.usp.br}
\affiliation{Instituto de F\'\i sica, Universidade de S\~ao Paulo\\
Caixa Postal 66318, 05314-970, S\~ao Paulo, SP, Brazil}


\begin{abstract}

We establish the equivalence between the continuum limit of the quantum spherical model with competing interactions, which is relevant
to the investigation of Lifshitz points, and the $O(N)$ nonlinear sigma model with the addition of higher order spatial derivative operators,
which breaks the Lorentz symmetry and is known as Lifshitz-type (or anisotropic) nonlinear sigma model.
In the context of the $1/N$ expansion, we also discuss the renormalization properties of this nonlinear sigma model
and find the nontrivial fixed points of the $\beta$-functions in various dimensions, which turn out to be connected with the existence of phase transitions in the quantum spherical model.

\end{abstract}

\maketitle

\section{Introduction}

The connection between quantum field theory and critical phenomena has been a prosperous subject over the years and
in fact it has produced an interesting confluence of ideas and unification of several concepts, providing new and relevant insights.
One of best examples of this relation is the renormalization group \cite{Wilson}, that is an essential framework
to deal with systems involving a large number of degrees of freedom.

This work is dedicated to the study of the equivalence
between the quantum spherical model with competing
interactions and the $O(N)$ nonlinear sigma model with addition of higher spatial derivative operators in the limit
of $N$ tending to infinity. We explore several properties at both sides of this connection,
as the existence of Lifshitz points and the critical behavior of the competitive quantum spherical model, as well as renormalization and
renormalization group aspects of the anisotropic nonlinear sigma model.
This relationship generalizes the equivalence between the quantum spherical model with short-range interactions
and the relativistic nonlinear sigma model~\cite{Vojta1}.

The classical spherical model was introduced by Berlin and Kac in 1952 \cite{Berlin}. It is a representative example of a soluble statistical model that exhibits a nontrivial critical behavior. Due to these characteristics, it became an excellent laboratory to investigate several aspects of phase transitions and critical phenomena \cite{Joyce}.
The quantum versions of the spherical model \cite{Vojta1,Obermair,Henkel1,Nieuwenhuizen} share with the classical counterpart essentially the
same good characteristics and are then instrumental in the study of critical phenomena at very low temperatures (eventually zero) as
quantum phase transitions~\cite{Sachdev}.

One aspect that plays a central role throughout this work is the so called Lifshitz point.
A Lifshitz point in a given phase diagram is the meeting point between ordered, disordered, and
modulated phases \cite{9b}. This kind of structure is of great interest since it
occurs in several systems as magnetic compounds, liquid crystals, and polymers \cite{Lubensky,Diehl}.

An essential requirement for describing modulated structures in the phase diagram is the presence of competing interactions
favoring different orderings \cite{Selke,Leite1,Leite2}. A well known example is the Axial-Next-Nearest-Neighbor-Ising-model, designated as
ANNNI-model \cite{Selke1}, with
ferromagnetic interactions between first neighbors along all directions and
antiferromagnetic interactions between second neighbors along one specific direction, originating the competition ferro/antiferro in that direction.
This model exhibits modulated, disordered, and ordered phases, meeting at a
Lifshitz point.

Of special interest for the developments of this work is the case in which
the competing interactions are extended to $m\leq d$ directions, where $d$ is the dimension of the lattice.
We will consider the quantum spherical model with competing interactions along $m=d$ directions plus a form of diagonal interactions that will be
specified shortly. Some studies with classical and quantum versions of the spherical model with competing interactions were performed
by a number of investigations \cite{Kalok,Pisani,Dutta,Bienzobaz}.

From the field theory perspective, the relativistic nonlinear sigma model has a long history \cite{Levy},
constituting an important prototype
both from the theoretical point of view as well as in phenomenological applications.
It is renormalizable in two dimensions in the perturbative expansion \cite{Brezin} and in two
and three spacetime dimensions in the context of the $1/N$ expansion \cite{Arefeva,Rosenstein},
exhibiting interesting properties as dynamical mass generation and asymptotic freedom \cite{Polyakov} .

As we shall see, by taking the continuum limit of the quantum spherical model
with competing interactions, we are naturally led to a nonlinear sigma model with the
presence of higher spatial derivative operators. Despite of the obvious Lorentz symmetry breaking,
this model has better ultraviolet behavior as compared with the relativistic one and
so opens the possibility to obtain renormalizable sigma models in higher dimensions.

From a pure field theory perspective, the possibility of constructing unitary theories with
a better ultraviolet behavior by adding higher spatial derivative operators has drawn much attention recently \cite{Anselmi1,Anselmi,Chen,Alexandre}.
For example, in this framework it has been argued to be possible to construct
a quantum gravity theory that is power counting renormalizable \cite{Horava1}. In this context, the Lorentz symmetry would arise as a
low energy manifestation. In spite of the plausibility of such idea, renormalization group studies of these theories
have shown that this is a delicate point, in general depending on specific fine tunes \cite{Iengo,Gomes}.

Our work is organized as follows. In section \ref{section2}, we discuss
aspects of the quantum spherical model with competing interactions, as the determination of certain relations between the parameters of the model,
corresponding to different phases separated by the Lifshitz point, as well as the continuum limit.
Section \ref{section3} is focused on the equivalence between the spherical model with competing interactions and the anisotropic
nonlinear sigma model by considering the large-$N$ effective action.
In section \ref{four}, we study the quantum critical behavior of the competing spherical model and determine the critical dimensions.
Section \ref{section5} is dedicated to the study of renormalization of the anisotropic nonlinear sigma models.
A summary and additional comments are presented in section \ref{Summary}. There is also an Appendix
in which we illustrate the application of the subtraction scheme to a divergent Green function
involved in the renormalization procedure.


\section{The Quantum Spherical Model with competing interactions}\label{section2}

We start this section by outlining some basic facts about the
classical spherical model as well as its quantum version.
The classical Hamiltonian is defined by
\begin{equation}
\mathcal{H}_c=-\frac12\sum_{{\bf r},{\bf r}'}J_{{\bf r},{\bf r}'}S_{\bf r}S_{{\bf r}'}-h\sum_{\bf r}S_{\bf r},
\label{1.1}
\end{equation}
where ${\bf r}$ and ${\bf r}'$ are lattice vectors, $\left\{S_{\bf r}\right\}$ is a set of continuous spin variables
on a $d$-dimensional hypercubic lattice with periodic boundary conditions,
$J_{{\bf r},{\bf r}^{\prime}}$ is the interaction energy that depends only on the distance between the sites
${\bf r}$ and ${\bf r}'$, $J_{{\bf r},{\bf r}'}\equiv J(|{\bf r}-{\bf r}'|)$, and $h$ is an external field.
The spin variables are subject to the spherical constraint
\begin{equation}
\sum_{\bf r}S_{\bf r}^2=N,
\label{1.2}
\end{equation}
with $N$ being the total number of lattice sites.

The quantum version of this model can be obtained as follows \cite{Vojta1,Obermair}.
We first add to the Hamiltonian a kinetic term involving the conjugated momentum to $S_{\bf r}$,
denoted by $P_{\bf r}$, such that
\begin{equation}
\mathcal{H}=\frac{g}{2}\sum_{\bf r}P_{\bf r}^2-\frac12\sum_{{\bf r},{\bf r}'}J_{{\bf r},{\bf r}'}S_{\bf r}S_{{\bf r}'}-h\sum_{\bf r}S_{\bf r}.
\label{1.3}
\end{equation}
The parameter $g$ plays the role of a quantum coupling. By assuming the
commutation relations
\begin{equation}
[S_{\bf r}, S_{{\bf r}^{\prime}}]=0, ~~~[P_{\bf r}, P_{{\bf r}^{\prime}}]=0~~~\text{and}~~~
[S_{\bf r}, P_{{\bf r}^{\prime}}]=i\delta_{{\bf r}, {\bf r}^{\prime}},
\label{1.4}
\end{equation}
we obtain the quantum version.
In this approach, it is easier to implement the so called mean spherical model, which means that the constraint must be enforced
as a thermal average, $\sum_{\bf r}\langle S_{\bf r}^2\rangle=N$.
Alternatively, we may simply consider the variables as classical ones and proceed with the quantization by means of a path integral.
This last approach is more appropriate to impose the strict spherical constraint (\ref{1.2}) and furthermore
makes clear the connection with the nonlinear sigma model.
We will return to it in Section \ref{section3} when discussing the equivalence between models.

Now let us consider a particular form for the exchange energy $J_{{\bf r},{\bf r}'}$, which involves competing interactions.
We assume ferromagnetic interactions between first neighbors and antiferromagnetic interactions between second
neighbors and also between diagonal neighbors belonging to the same plane.
As mentioned in the Introduction, this is the generalization of the ANNNI-interactions along all directions
plus the diagonal interactions. In fact, this is the isotropic case in the
sense that the interactions are equally distributed along all directions.
For the case of the spherical version it is also denoted as
ANNNS-model (see for example \cite{Henkel}).

\begin{figure}[!h]
\centering
\includegraphics[scale=0.6]{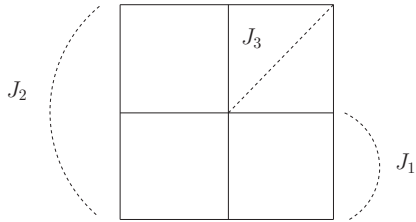}
\caption{Competing interactions in the square lattice.}
\label{CompeticaoFig}
\end{figure}

For concreteness, we shall consider the spherical model on a two-dimensional square lattice,
although we will generalize the analysis to arbitrary higher dimensions.
Hereafter, we consider the system in the absence of the external field, $h=0$.
The Hamiltonian can be written as
\begin{equation}
\mathcal{H}=\frac{g}{2}\sum_{\bf r}P_{\bf r}^2-J_1\sum_{<{\bf r},{\bf r}'>}S_{\bf r}S_{{\bf r}'}-J_2\sum_{\ll{\bf r},{\bf r}'\gg}S_{\bf r}S_{{\bf r}'}-
J_3\sum_{\prec{\bf r},{\bf r}'\succ}S_{\bf r}S_{{\bf r}'},
\label{1.5}
\end{equation}
with $J_1>0$ favoring the ferromagnetic ordering, $J_2<0$ and $J_3<0$ favoring the antiferromagnetic.
The symbols $<>$ and $\ll\gg$ indicate a sum restricted to the first and second neighbors (along the Cartesian axes) respectively, whereas $\prec\,\succ$ means
a sum restricted to diagonal neighbors. The geometric illustration is shown in Fig. \ref{CompeticaoFig}.
The usual isotropic ANNNS case corresponds to $J_3=0$ and its quantum version was analyzed in \cite{Bienzobaz}. The diagonal interaction $J_3$ has an important role in
the continuum limit as we shall see.

We may determine certain relations between the parameters $J_1,J_2$, and $J_3$, corresponding to different phases in an appropriate phase diagram,
by analyzing the maximum of the Fourier transformation of the interaction energy,
\begin{equation}
J({\bf q})=\sum_{\bf h}J(|{\bf h}|)e^{i {\bf q}\cdot{\bf h}},~~~\text{with}~~~{\bf h}={\bf r}-{\bf r}^{\prime}.
\label{1.6}
\end{equation}
For the interactions in (\ref{1.5}), in the two-dimensional case, $J({\bf q})$ becomes
\begin{equation}
J({\bf q})=2J_1[\cos(q_x)+\cos(q_y)]+2J_2[\cos(2q_x)+\cos(2q_y)]+4J_3\cos(q_x)\cos(q_y).
\label{1.7}
\end{equation}
Looking for the maximum values of $J(\bf q)$ and the corresponding values for ${\bf q}$, designated by ${\bf q}^c$, we find
\begin{equation}
q_x^c=q_y^c=0,~~~\text{for}~~~p\leq1/4
\label{1.8}
\end{equation}
and
\begin{equation}
q_x^c=q_y^c=\cos^{-1}\left(\frac{-J_1}{4J_2+2J_3}\right),~~~\text{for}~~~p>1/4,
\label{1.9}
\end{equation}
where $p\equiv \frac{-(J_2+J_3/2)}{J_1}$ and $J_3\neq 2J_2$.

The case $J_3= 2J_2$ must be treated separately, and gives
\begin{equation}
q_x^c=q_y^c=0, ~~~\text{for} ~~~\tilde{p}\leq1/4,
\label{1.10}
\end{equation}
with $\tilde{p}\equiv-2J_2/J_1$. The solution for values of $q_x^c$ and $q_y^c$ different from zero
does not completely fix its values. Instead, it gives a restriction over the sum
\begin{equation}
\cos{q_x^c}+\cos{q_y^c}=\frac{-J_1}{4J_2}, ~~~\text{for}~~~\tilde{p}>1/4.
\label{1.11}
\end{equation}
Thus we have freedom to choose the value for one of these parameters, say $q_x^c$, and the above condition determine $q_y^c$
as a function of $q_x^c$. This choice does not affect the critical behavior of the model, as may be seen from the analysis in the Section \ref{four}.
For simplicity, let us consider the symmetric case where $q_x^c=q_y^c$, which yields
\begin{equation}
\cos{q_x^c}=\cos{q_y^c}=\frac{-J_1}{8J_2}.
\label{1.12}
\end{equation}
With this choice, we may unify the results for $J_3\neq 2 J_2$ and $J_3= 2 J_2$
simply by considering the relations (\ref{1.8}) and (\ref{1.9}) for all values of $J_2$ and $J_3$.
We observe that the point $p=1/4$ is a divider separating regions with different critical values of ${\bf q}$.
It is called a Lifshitz point, corresponding to the meeting point between ordered, disordered and the
modulated phases.

In order to further explore the properties of the system at the Lifshitz point,
let us expand $J({\bf q})$ around the critical value ${\bf q}^c=(0,0)$,
\begin{equation}
J({\bf q})=4(J_1+J_2+J_3)-(J_1+4J_2+2J_3){\bf q}^2+\frac{1}{4!}(2J_1+32J_2+4J_3)(q_x^4+q_y^4)+J_3 q_x^2q_y^2+\cdots.
\label{1.13}
\end{equation}
Observe that at the Lifshitz point, $p=1/4$, the coefficient of the quadratic term vanishes, $J_1+4J_2+2J_3=0$,
and the term of fourth order becomes important. The consequence of this is a different behavior
between space and time in the system under
a scaling transformation with parameter $\theta$, i.e., $t\rightarrow \theta t$ whereas ${\bf r}\rightarrow \theta^2 {\bf r}$.

By taking the Lifshitz point, we may eliminate $J_1$ in the expression (\ref{1.13}) by means of $J_1=-4J_2-2J_3$, such that it reduces to
\begin{equation}
J({\bf q})=-4(3J_2+J_3)+J_2(q_x^4+q_y^4)+J_3 q_x^2q_y^2+\cdots.
\label{1.14}
\end{equation}
We immediately note that when $J_3=2J_2$ this expression can be written
in a rotational invariant way in terms of the modulus of ${\bf q}$,
\begin{equation}
J({\bf q})=-20J_2+J_2|{\bf q}|^4+\cdots.
\label{1.15}
\end{equation}
Incidently, this is exactly the point that was treated separately
culminating with conditions (\ref{1.10}) and (\ref{1.11}). The issue of rotational
invariance will be important to obtain a field theory in the continuum limit.
The generalization for higher dimensional lattices
is straightforward and will be discussed from now on.

The generalization of the equation ({\ref{1.7}) for an arbitrary $d$-dimensional lattice is given by
\begin{eqnarray}
J({\bf q})=2 J_1\sum_{i=1}^d\cos q_i+2J_2\sum_{i=1}^d\cos 2q_i
+ 4J_3\sum_{i<j}^d\cos q_i\cos q_j,
\label{1.16}
\end{eqnarray}
and it is straightforward to find the maximum values of $J(\bf q)$. As in the two-dimensional case,
for $p\leq 1/4$, we have ${\bf q}^c=(0,0,\ldots,0)$. For $p>1/4$, with the same assumption that led us to the equation (\ref{1.12}),
namely, $q_1^c=q_2^c=\cdots= q_d^c$, we have
\begin{eqnarray}
\cos q^c_i=\frac{-J_1}{4\left[ J_2+\frac12(d-1)J_3\right]},
\label{1.17}
\end{eqnarray}
with $i=1,...,d$ and $p\equiv \frac{-[ J_2+\frac12(d-1)J_3]}{J_1}$. Of course that, when $d=2$ we recover the equation (\ref{1.9})
as well as the corresponding relation for $p$. The expansion of $J({\bf q})$ around ${\bf q}^c=(0,0,\ldots,0)$ takes the form
\begin{eqnarray}
J({\bf q})&=& 2d[J_1+J_2+(d-1)J_3]-[J_1+4J_2+2(d-1)J_3]|{\bf q}|^2\nonumber\\&+&\frac{1}{12}[J_1+16J_2+2(d-1)J_3]\sum_i^d q_i^4+J_3\sum_{i<j}q_i^2q_j^2+\cdots.
\label{1.18}
\end{eqnarray}
By taking the Lifshitz point and $J_3=2J_2$ it reduces to
\begin{equation}
J({\bf q})=-2d(1+2d)J_2+J_2|{\bf q}|^4+\cdots.
\label{1.19}
\end{equation}

\subsection{Continuum Limit}

We now investigate the connection between the quantum spherical model with competing interactions
and a special form of the nonlinear sigma model.
A natural way to investigate this connection is just by taking the continuum limit of the lattice.
In this situation, the lattice is replaced by a continuous structure giving rise to
an underlying field theory and it would be interesting to recognize this field theory.
In the above discussion, the spacing between the sites was set $a\equiv 1$, but now we have to restore it, in order
to take $a\rightarrow 0$. The analysis below show us precisely the effect of each interaction
between neighbors on the corresponding field theory.

For our purpose here it is convenient to consider the Lagrangian, obtained from a Legendre transformation of (\ref{1.5}),
\begin{equation}
L=\frac{1}{2g}\sum_{\bf r}\left(\frac{dS_{\bf r}}{dt}\right)^2+J_1\sum_{<{\bf r},{\bf r}'>}S_{\bf r}S_{{\bf r}'}+J_2\sum_{\ll{\bf r},{\bf r}'\gg}S_{\bf r}S_{{\bf r}'}+J_3\sum_{\prec{\bf r},{\bf r}'\succ}S_{\bf r}S_{{\bf r}'}.
\label{2.1}
\end{equation}
Now we write this expression in a more explicit way, by detailing the interactions between the neighbors over the two-dimensional lattice,
\begin{eqnarray}
L&=&\frac{1}{2g}\sum_{x,y}\left(\frac{dS_{x,y}}{dt}\right)^2+J_1\sum_{x,y}(S_{x,y}S_{x+a,y}+S_{x,y}S_{x,y+a})+
J_2\sum_{x,y}(S_{x,y}S_{x+2a,y}+S_{x,y}S_{x,y+2a})\nonumber\\&+&J_3\sum_{x,y}(S_{x,y}S_{x+a,y+a}+S_{x+a,y}S_{x,y+a}),
\label{2.2}
\end{eqnarray}
and also the spherical constraint,
\begin{eqnarray}
\sum_{{x,y}}S_{xy}^2=N.
\label{2.3}
\end{eqnarray}

In the continuum limit of the lattice, $a\rightarrow 0$, the variables $S_{\bf r}$ become functions of the continuous
position variable ${\bf r}$, $S_{\bf r}(t)\rightarrow S({\bf r},t)$,
the sums are replaced by integrals according to $a^d\sum_{\bf r}\rightarrow \int d^dr$ and
the interactions are identified with derivatives.
For the first neighbors interactions, the term $S_{x+a,y}S_{x,y}$, for example, can be written as
\begin{equation}
S_{x+a,y}S_{x,y}\sim S_{x,y}^2-\frac12~ a^2\left(\frac{\partial S_{x,y}}{\partial x}\right)^2.
\label{2.4}
\end{equation}
In this expression we took advantage of the sum over $x$ and $y$ that is present in the Lagrangian,
which implies equalities as $\sum_{xy}S^2_{x+a,y}=\sum_{xy}S^2_{x,y}$.
Similarly, the interaction between second neighbors, $S_{x+2a,y}S_{x,y}$, becomes
\begin{equation}
S_{x+2a,y}S_{x,y}\sim S_{x,y}^2-2a^2\left(\frac{\partial S_{x,y}}{\partial x}\right)^2+
\frac12 a^4\left(\frac{\partial^2 S_{x,y}}{\partial x^2}\right)^2.
\label{2.5}
\end{equation}
The interactions between first and second neighbors along the $y$ direction are analogous the equations
(\ref{2.4}) and (\ref{2.5}), respectively.
Finally, the diagonal interactions are written as
\begin{equation}
S_{x+a,y+a}S_{x,y}+S_{x+a,y}S_{x,y+a}\sim 2S_{x,y}^2-a^2 \left(\frac{\partial S_{x,y}}{\partial x}\right)^2-
a^2\left(\frac{\partial S_{x,y}}{\partial y}\right)^2+\frac12 a^4 \left(\frac{\partial^2 S_{x,y}}{\partial x\partial y}\right)^2.
\label{2.6}
\end{equation}
In equations (\ref{2.5}) and (\ref{2.6}) we also used the fact that there is a sum over $x$ and $y$.
By summing all contributions, the Lagrangian (\ref{2.2}) takes the form
\begin{eqnarray}
L&=&\int \frac{d^2r}{a^2} \left\{\frac{1}{2g}\left(\frac{\partial S}{\partial t}\right)^2+2(J_1+J_2+J_3)S^2-\frac12a^2(J_1+4J_2+2J_3)(\nabla S)^2
\right.\nonumber\\&+&\left.
\frac12 a^4 J_2 \left[\left(\frac{\partial^2 S}{\partial x^2}\right)^2+\left(\frac{\partial^2 S}{\partial y^2}\right)^2
+\frac{J_3}{J_2} \left(\frac{\partial^2 S}{\partial x\partial y}\right)^2\right]\right\},
\label{2.7}
\end{eqnarray}
where we are omitting the spacetime dependence, $S\equiv S(x,y,t)$.
The spherical constraint in this limit becomes
\begin{equation}
\int d^2r\, S^2=Na^2\equiv \text{fixed}.
\label{2.8}
\end{equation}

The Lifshitz point is characterized by the vanishing of the coefficient of the term $(\nabla S)^2$ in (\ref{2.7}),
i.e., when $J_1+4J_2+2J_3=0$, in accordance with the discussion below equation (\ref{1.13}).
Its special feature, as already discussed, is the anisotropic scaling between space and time, with a dynamical critical exponent $z=2$.
On the other hand, when  $J_2=J_3=0$ (without competing interactions), the Lagrangian (\ref{2.7}) exhibits the relativistic symmetry, corresponding to $z=1$.

From this analysis it is clear that the inclusion of interactions between more distant neighbors
is equivalent, in the field theory side (continuum limit), to consider spatial derivative terms of higher order, which may give rise
to arbitrary values for the dynamical critical exponent~$z$.

For field theoretical purposes, it is desirable to have a rotational invariant Lagrangian, which requires that the last line
of (\ref{2.7}) must be recognized as $(\nabla^2S)^2$, where $\nabla^2\equiv\partial_i\partial_i$
(with the sum convention over repeated indices).
This is obtained when $J_3=2J_2$, and we get
\begin{eqnarray}
L=\int \frac{d^2r}{a^2} \left[\frac{1}{2g}\left(\frac{\partial S}{\partial t}\right)^2+2(J_1+3J_2)S^2-\frac12a^2(J_1+ 8J_2)(\nabla S)^2
+\frac12 a^4 J_2 (\nabla^2S)^2\right],
\label{2.9}
\end{eqnarray}
up to surface terms that vanish due to the periodic boundary conditions.
Notice that the lattice spacing  $a$ can be eliminated through the rescaling  $x^i\rightarrow a x^i$.
This Lagrangian shows the structure of spatial derivatives in the field theory arising as the continuum limit of a model
with competing interactions.

The generalization of this result for arbitrary dimensions $d$ is straightforward and we just write the result for the
Lagrangian in the continuum limit, assuming $J_3=2J_2$,
\begin{eqnarray}
L=\int \frac{d^dr}{a^d} \left\{\frac12\left(\frac{\partial S}{\partial t}\right)^2+d[J_1+(2d -1)J_2]S^2
-\frac12a^2[J_1+ 4 d J_2](\nabla S)^2
+\frac12 a^4 J_2 (\nabla^2S)^2\right\},
\label{2.10}
\end{eqnarray}
whereas the constraint is
\begin{equation}
\int d^dr\, S^2=Na^d\equiv \text{fixed}.
\label{2.11}
\end{equation}

The development presented along this section revealed the characteristics of the field theory underlying the spherical model with
competing interactions. In the following we formulate and proceed with the quantization of the nonlinear sigma model with the addition of a
higher spatial derivative term in order to investigate the equivalence with the competing quantum spherical model.


\section{the quantum spherical and the Nonlinear sigma models}\label{section3}

In this section we establish, at the quantum level, the equivalence between the strictly quantum spherical model with competing interactions
and the nonlinear sigma model with a higher spatial derivative operator, in the limit of the number of fields tending to infinite.
This equivalence holds in the sense that both partition functions and, consequently, the quantities that follow from them coincide.
Our strategy is to consider the effective action of the nonlinear sigma model in the context of the
$1/N$ expansion and then take the limit $N\rightarrow\infty$.

As we mentioned in the section \ref{section2}, the strictly
quantum spherical model can be obtained by the functional integration \cite{Gomes2}
\begin{equation}
Z=\int \mathcal{D}S_{\bf r}\,\delta\Big{(}\sum_{\bf r}S_{\bf r}^2-N\Big{)}e^{-\int_{0}^{\beta}d\tau L_E},
\label{3.1}
\end{equation}
with the Euclidean Lagrangian,
\begin{equation}
L_E=\frac{1}{2g} \sum_{\bf r}\left( \frac{\partial S_{\bf r}(\tau)}{\partial\tau}\right)^2
-\frac12\sum_{{\bf r},{\bf r}^{\prime}}J_{{\bf r},{\bf r}^{\prime}}S_{{\bf r}}S_{{\bf r}^{\prime}},
\label{3.2}
\end{equation}
where $\tau=it$, $\tau\in[0,\beta]$ ($\beta$ is the inverse of the temperature), and the variables $S_{\bf r}(\tau)$ satisfying
the periodic condition in the imaginary time $S_{\bf r}(0)=S_{\bf r}(\beta)$.
The functional integration measure $\mathcal{D}S_{\bf r}$ symbolically
stands for the product over all sites, $\prod_{\bf r}\mathcal{D}S_{\bf r}$.

By employing the saddle point method, which is exact in the thermodynamic limit, we obtain the
saddle point condition
\begin{equation}
1-\frac{1}{N}\sum_{\bf q}\frac{g}{2\omega_{\bf q}}\coth\left(\frac{\beta\omega_{\bf q}}{2}\right)=0,
\label{3.3}
\end{equation}
with $\omega_{\bf q}^2\equiv 2g(\mu-J({\bf q})/2)$, and $\mu$ being the saddle point value of the auxiliary field that
implements the constraint (Lagrange multiplier).
In the thermodynamic limit the sum over the momentum ${\bf q}$ must
be understood as an integral, $\frac{1}{N}\sum_{\bf q}\rightarrow \int d^d q$.
As we saw in Section \ref{section2}, the expansion of
$J({\bf q})$ around its critical value has the structure
\begin{equation}
J({\bf q})=A_0+A_1|{\bf q}|^2+A_3\sum_i^d q_i^4+A_4\sum_{i<j}q_i^2q_j^2+\cdots,
\label{3.4}
\end{equation}
where the coefficients $A_i$ depend on the interaction parameters and on the dimension $d$ of the lattice, $A_i\equiv A_i(J_1,J_2,J_3,d)$,
that can be obtained from equation (\ref{1.18}).
At the Lifshitz point, where $A_1=0$, and with
the special relation between the parameters $J_3=2J_2$, $J({\bf q})$ reduces to
\begin{equation}
J({\bf q})=\widetilde{A}_0+\widetilde{A}_1|{\bf q}|^4+\cdots.
\label{3.5}
\end{equation}
As we shall discuss in the next section, the critical behavior of the system can be studied by considering the
saddle point condition (\ref{3.3}) near the critical point, with the above forms for $J({\bf q})$.

Now let us consider the nonlinear sigma model. The $O(N)$ anisotropic ($z=2$) nonlinear sigma model
involves $N$ scalar fields,  $\varphi_a$, $a=1,...,N$, with the Lagrangian including a higher spatial derivative operator term,
\begin{equation}
\mathcal{L}=\frac12\partial_{0}\varphi\partial_{0}\varphi-\frac{a_1^2}{2}\partial_{i}\varphi\partial_{i}\varphi-
\frac{a_2^2}{2}\Delta\varphi\Delta\varphi,
\label{3.6}
\end{equation}
where $\Delta\equiv\nabla^2$, and we are omitting the $O(N)$ index $a$. The fields $\varphi_a$ are subjected to the constraint
\begin{equation}
\varphi_a^2\equiv\varphi^2=\frac{N}{2g},
\label{3.7}
\end{equation}
with $g$ being the coupling constant. The relativistic situation corresponds to $a_2=0$ and $a_1\neq 0$, whereas
the analogous field theory Lifshitz point corresponds to the opposite case, $a_1=0$ and $a_2\neq 0$.
As it will be discussed later, both terms may be necessary for the renormalization in the anisotropic case.
Due to the constraint we may add a mass term
($\sim\varphi^2=\text{constant}$) to the Lagrangian without modifying the physical content of the theory,
\begin{equation}
\mathcal{L}=\frac12\partial_{0}\varphi\partial_{0}\varphi-\frac{a_1^2}{2}\partial_{i}\varphi\partial_{i}\varphi-
\frac{a_2^2}{2}\Delta\varphi\Delta\varphi-\frac{m^4}{2}\varphi^2.
\label{3.8}
\end{equation}

The procedure for the determination of the effective action, in the context of the $1/N$ expansion,
may be outlined as follows. The constraint is implemented by means of a delta function that is written
in terms of an integral over some auxiliary field, say $\sigma$, playing the role of a Lagrange multiplier.
With this, we may perform the integration over the fields $\varphi_a$ and then obtain an
effective action in terms of $\sigma$. The effective action has the structure of an  $1/N$
expansion,
\begin{equation}
S_{eff}=N^{1/2}S_1+N^0S_2+N^{-1/2}S_3+\cdots,
\label{3.9}
\end{equation}
with $S_n$ being the contribution for the effective action of the referred order in $1/N$, and
$n$ indicating the corresponding power of the auxiliary field $\sigma$.
To make sense of the expansion (\ref{3.9}) as $N\to\infty$, it is necessary the vanishing of the term $S_1$, associated with the
positive power of $N$. This will lead us to the gap equation, which in the Euclidean space reads
\begin{equation}
\frac{1}{2g}-\int \frac{d^{d+1}q}{(2\pi)^{d+1}}\frac{1}{q_0^2+a_1^2{\bf q}^2+a_2^2({\bf q}^2)^2+m^4}=0.
\label{3.10}
\end{equation}
This equation is similar to the saddle point condition (\ref{3.3}) in the thermodynamic limit.
In fact, by considering the system at finite temperature, we need to take into account that the
integral over momentum (zero component) is replaced by a sum over the Matsubara frequencies, such that
\begin{equation}
\int\frac{d^{d+1}q}{(2\pi)^{d+1}}\rightarrow\frac{1}{\beta}\sum_n\int\frac{d^{d}q}{(2\pi)^{d}},
\label{3.10a}
\end{equation}
and $q_0\rightarrow \omega_n=\frac{2\pi n}{\beta}$, with $n\in \text{Z}$. The sum over $n$ can be evaluated according to
\begin{equation}
\sum_{n=-\infty}^{\infty}\frac{1}{n^2+y^2}=\frac{\pi}{y}\,\coth(\pi y),~~~y>0,
\label{3.11}
\end{equation}
which enable us to get the final expression
\begin{equation}
\frac{1}{2g}-\int \frac{d^{d}q}{(2\pi)^{d}}\frac{1}{2\omega_q}\coth\left(\frac{\beta\omega_q}{2}\right)=0,
\label{3.12}
\end{equation}
with $\omega_q\equiv\sqrt{a_1^2{\bf q}^2+a_2^2({\bf q}^2)^2+m^4}$. This equation must be compared with (\ref{3.3}).
 With an appropriate identification between the parameters
of the quantum spherical model and of the nonlinear sigma model we may establish the following equivalences.
First, outside the Lifshitz point, $J({\bf q})$ is dominated by a quadratic term in the momenta, as can be seen from (\ref{3.4}).
This corresponds to the relativistic situation, where $a_1\neq 0$ and $a_2=0$.
Second, at the Lifshitz point, the quadratic term in $J({\bf q})$ vanishes and $J({\bf q})$ has the form (\ref{3.5}),
corresponding to the choice of parameters $a_1=0$ and $a_2\neq 0$, which is the nonrelativistic nonlinear sigma model with
the presence of a higher spatial derivative term.

An important observation concerns to the integration limits.
In the integral in equation (\ref{3.12}) they do not have any restriction, whereas in (\ref{3.3}) they belong to the first Brillouin zone.
Actually, the equivalence is achieved in the continuum limit, with the lattice spacing $a\rightarrow 0$.
In the case of the spherical model, we were considering unitary spacing, such that it did not appear explicitly.
By restoring its dependence, the first Brillouin zone, that for a hypercubic lattice is delimited by $[-\pi/a,\pi/a]$ for each
momentum component, will extend to the infinity. The last step in order to establish the complete equivalence
is by taking the limit $N\rightarrow\infty$. This means that in the effective action (\ref{3.9}) only the $S_2$ term will
contribute. This is exactly the Gaussian approximation for the $\sigma$ integration, equivalent to the saddle point
method.

\section{Quantum Critical Behavior}\label{four}

In this section we will discuss the critical behavior of the quantum spherical model with competing interactions
in order to verify the existence of phase transitions and then
determine the lower and upper critical dimensions. We can identify
the dimensions in which the system exhibits trivial (mean-field) and nontrivial critical behaviors, or even there is not a phase transition. We will not perform the analysis of the behavior of thermodynamic quantities
nor will calculate critical exponents.  The results obtained here will be
contrasted with the $\beta$-function of the renormalization group of the nonlinear sigma model
and are related to the existence of trivial and nontrivial fixed points.

In the study of critical behavior we essentially need to analyze the convergence properties of the integral in the
saddle point condition (\ref{3.3}) and the dependence with the parameters $\mu$, $g$, and the temperature $T$.
We will consider two cases separately according to the values of the parameter $p$, i.e., $p\neq 1/4$ and $p=1/4$, because the different forms for the expansion of $J({\bf q})$ in each of these situations will lead different convergence properties.
As we are interested in quantum phase transitions, we study only the transitions that occur at
zero temperature.

The critical behavior can be obtained by analyzing the equation (\ref{3.3}) near the critical point.
Actually, we first consider it exactly at the critical point, where the parameters assume the critical values $\mu_c$ and $g_c$.
Next, we consider this expression near the critical point. In this case, we expand $J({\bf q})$ around the critical point and then subtract it from the equation at the critical point.
The difference between them enable us to relate the chemical potential $\mu$ in terms of the
distance from the quantum critical point, that we defined as $\tau\equiv (g-g_c)/g_c$.
The critical value that maximizes the interaction energy $J({\bf q})$ depends on the value of parameter $p$,
as discussed in section II.

Analyzing the convergence of equation (\ref{3.3}) we see that for $p\neq 1/4$ the sum converges if $d>1$, defining
the lower critical dimension, $d_l=1$. In this case, we have
\begin{equation}
(\mu-\mu_c)\sim\left\{
\begin{array}{cc}
\tau,&~~~~ d>3\\
\frac{\tau}{\ln \tau},&~~~~d=3\\
\tau^2,&~~~~d=2
\end{array}
\right.,
\label{4.1}
\end{equation}
with $\tau=(g-g_c)/g_c$. There is no phase transition for $d=1$. For $d=2$ the system exhibits
a critical point with nontrivial critical behavior, for $d>3$ we have mean-field critical behavior, and $d=3$ is the threshold dimension
between these two behaviors involving logarithmic corrections to the mean-field, defined as the upper critical dimension.

At the Lifshitz point, $p=1/4$, the sum in (\ref{3.3}) converges if $d>2$, which defines the lower critical dimension, $d_l=2$.
So, for $d=2$ the system does not exhibit a phase transition. For other dimensions, we obtain
\begin{eqnarray}
(\mu-\mu_c)\sim\left\{
\begin{array}{cc}
\tau,&~~~ d>6\\
\vspace{0.1cm}
\frac{\tau}{\ln \tau},&~~~d=6\\
{\tau}^{3/2},&~~~d=5\\
\tau^2,&~~~d=4\\
\tau^{5/2},&~~~d=3
\end{array}
\right..
\label{4.2}
\end{eqnarray}
In this case, for $d=3,4$, and 5 the
system exhibits nontrivial critical behaviors. For $d>6$ we obtain a mean-field behavior, and $d=6$ is the threshold dimension
(upper critical dimension) between these behaviors with logarithm corrections to the mean-field.

We may compare the above results with that of reference \cite{Bienzobaz} by means of an appropriate redefinition of the involved parameters.
In the mentioned reference it was analyzed the quantum spherical model with ferromagnetic interactions $J_1>0$ between first neighbors along all directions
and antiferromagnetic interactions $J_2<0$ between second neighbors along $m\leq d$ directions, originating the competition.
By redefining a combination of antiferromagnetic interactions energies $J_2$ and $J_3$ as
$J_2+\frac12(d-1)J_3\rightarrow J_2$, we obtain exactly the results of \cite{Bienzobaz} in the case of competing interactions along all dimensions $m=d$ (isotropic case).

Finally, as we shall see later, the existence of nontrivial critical behavior is connected with the
existence of the nontrivial fixed points in the $\beta$-functions of the nonlinear sigma model.


\section{Generalized Anisotropic nonlinear sigma model}\label{section5}
In the previous section we have seen how the anisotropic nonlinear sigma model is related with the continuum limit of the spherical model with competing interactions. Now we will discuss some aspects of the former model  by considering a generalization  of the Lagrangian (\ref{3.6}) for arbitrary values of $z$, namely
\begin{equation}
{\cal L}= \frac12\partial_0 \varphi_{a}\partial_0 \varphi_{a}-\frac12\sum_{s=1}^{z} a_{s}^2\partial_{i_1}\ldots\partial_{i_s}\varphi_{a}\,\partial_{i_1}\ldots\partial_{i_s}\varphi_{a} -\frac{\sigma}{\sqrt{2N}}
\left(\varphi^2 -\frac{N}{2g}\right).
\label{5.1}
\end{equation}
Classically, the field $ \varphi_{a}, \, i=a,\ldots,N$, must satisfy the equation of motion
\begin{equation}
\left[\partial_{0}^{2}+\sum_{s=1}^{z}a_{s}^2(-1)^{s}(\triangle)^{s}+\sqrt{\frac2N}\sigma\right ]\varphi_{a}=0
\label{5.2}
\end{equation}
and the constraint $ \varphi^{2 }= \frac{N}{2g} $.  At the quantum level, the presence of higher spatial derivatives terms
improves the ultraviolet behavior of Feynman amplitudes so enlarging the class of renormalizable
models.  In this context, we will analyze
the $1/N$ expansion for the nonlinear sigma model in various dimensions.

The Lagrangian (\ref{5.1}) furnishes the following propagators in the large-$ N $ limit:

1. Propagator for the $ \varphi_a $ field:
\begin{equation}
\Delta_{ab}(p)= \frac{i\delta_{ab}}{p_{0}^{2}-\sum_{s=1}^{z}a_{s}^2{\bf p}^{2s}-m^{2z}},
\label{5.3}
\end{equation}
where a mass term was included. Notice that the presence of a nonvanishing mass is essential to evade infrared divergences if $d=z$;

2. Propagator $\Delta_{\sigma}$ for the auxiliary field $ \sigma $:
\begin{equation}
-\Delta_{\sigma}^{-1}(p)=\int \frac{dk_{0}}{2\pi}\frac{d^{d}{ k}}{(2\pi)^d} \frac{i}{(k_{0}+p_{0})^{2}-\sum_{s=1}^{z}a_{s}^2({\bf {p+k}})^{2s}-m^{2z}}\frac{i}{k_{0}^{2}-{\sum_{s=1}^{z}a_{s}^2\bf {k}}^{2s}-m^{2z}}\, ,
\label{5.4}
\end{equation}
which is finite for $ 3z>d $. At the bordering situation, $ z=1$ and $ d=3 $, which corresponds to the Lorentz covariant setting, the integral is logarithmically divergent and the renormalizability requires the introduction of a vertex proportional to $ \sigma^{2} $, but this would turn the model
indistinguishable from a $\varphi^{4}$ theory destroying its geometric nature. In such condition, the model is therefore nonrenormalizable and can be at most treated as an effective low energy theory.  Let us therefore restrict ourselves to values of $ z $ and $ d $ such that $ 3z>d $. In that situation, for large momentum  the above integral behaves as $ {\bf p}^{d-3z} $. Thus, for a generic graph $ \gamma $ with $ L $ loops, $ n_{\varphi} $ and $ n_{\sigma} $ internal lines of the $ \varphi $ and $ \sigma $ fields we have the following degree of superficial divergence
\begin{equation}
d(\gamma)= (z+d)L-2zn_{\varphi}+(3z-d)n_{\sigma}=z+d+(d-z)n_{\varphi}+4zn_{\sigma}-(z+d)V,
\label{5.5}
\end{equation}
where $ V $ is the number of vertices of $ \gamma $. This could be further simplified using
\begin{equation}
2n_{\varphi}+N_{\varphi}= 2 V \qquad \mbox{and} \qquad 2n_{\sigma}+N_{\sigma}=V,
\label{5.6}
\end{equation}
with $ N_{\varphi} $ and $ N_{\sigma} $ being the number of external lines of the corresponding fields. This gives
\begin{equation}
d(\gamma)=z+d -\frac{(d-z)}{2}N_{\varphi}-2z N_{\sigma}.
\label{5.7}
\end{equation}
Notice that $ \frac{d-z}2 $ and $ 2z $ are precisely the canonical anisotropic dimension of $ \varphi $ and $ \sigma $ fields.

Renormalizability requires that graphs without external $ \varphi $ lines be finite. From the above expression and as remarked before, this will be possible only
if $ 3z>d $; it seems also convenient to impose $ z\leq d $ so that the divergence of individual graphs does
not increase with the number of external lines. Thus we shall have
\begin{equation}
3z>d\geq z,
\label{5.8}
\end{equation}
but we still have to discuss the divergences in the $ \varphi $ sector.  Graphs with $N_{\sigma}=0$ and $N_{\varphi}=2$ have degree of divergence $2 z$  so that the subtraction of this divergences will induce bilinear counterterms containing a number even of  derivatives ranging from 0 to $2z$.
Thus, by considering the simplest anisotropic situation, namely $z=2$ that from now on we assume,  we see from (\ref{5.8}) that
$d$ may vary from $2$ to $5$.  The model with $d=3$ may be useful in possible
phenomenological applications as it is  a four spacetime dimensional version, renormalizable as we shall prove, of the nonlinear sigma model.  The case with
$d=2$, is atypical since the basic field $\varphi$ is dimensionless and mass generation becomes crucial to eliminate infrared divergences; unless for the tadpole graphs, which are logarithmically divergent, all Feynman amplitudes are quartically divergent.

For all situations with $d=3,\,4$ or $5
$ the unrenormalized Lagrangian is given by
\begin{equation}
{\cal L}_{un}= \frac12 \partial_{0}\varphi \partial_{0}\varphi-\frac{a_{1}^2}2\partial_{i}\varphi \partial_{i}\varphi -\frac{a_{2}^2}{2}\partial_{i}\partial_{j}\varphi \partial_{i}\partial_{j}\varphi- \frac{m^{4}}2\varphi^{2}-\frac{\sigma}{\sqrt{2N}}
\left(\varphi^2 -\frac{N}{2g}\right),
\label{5.9}
\end{equation}
where, due to stability reasons, all parameters, $a_{1},\,a_{2},\, g$, and $m$ are taken to be nonnegatives. For $d=2$, as we shall argue shortly, the inclusion of quadrilinear derivative couplings is also necessary. A discussion which overlaps ours about power counting renormalizability
was done in \cite{Anselmi}.

Without loss of generality, we may assume that $\langle\sigma\rangle=0 $ as a nonzero value for this expectation value would merely change the coefficient of the mass term (we also assume $\langle \varphi_{i}\rangle =0$ so that rotational symmetry is not broken). Now, this condition implies that  the gap equation,
\begin{equation}
\frac{1}{2g}=\int \frac{dk_{0}}{2\pi}\frac{d^{d}k}{(2\pi)^{d}} \frac{i}{k_{0}^{2}-a_{1}^2{\bf k}^{2}-a_{2}^2({\bf k}^{2})^2-m^{4}},
\label{5.9a}
\end{equation}
\noindent
must be obeyed. In setting the Feynman rules the diagrams of the Fig. \ref{Forbiden} are forbidden as they have already been used
to construct the sigma propagator (\ref{5.4}) and above tadpole equation.
The integral in the above expression is divergent so that an, up to now unspecified, regularization is necessary.
From this relation we may determine the $\beta$-functions.

\begin{figure}[!h]
\centering
\includegraphics[scale=0.6]{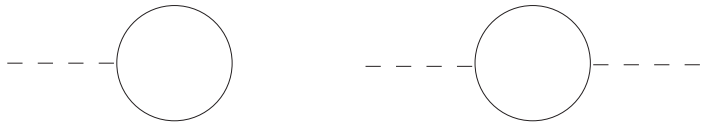}
\caption{Forbidden diagrams. The continuous and dashed lines represent the $\varphi$ and $\sigma$ field propagators, respectively.}
\label{Forbiden}
\end{figure}


\subsection{$\beta$-functions }

Let us calculate now the $\beta$-functions for the dimensions in which the model is renormalizable, $d=2,3,4$, and 5.
We have two distinct  situations: the case $d=2$, where the coupling constant is dimensionless;
and the cases $d=3,4$, and 5, where the coupling constant is a dimensionfull parameter.
As we shall see, the $\beta$-functions have different fixed points structure in these two situations, which is
related to the existence of quantum phase transitions.

We will consider the calculation of the $\beta$-functions at the Lifshitz point, i.e., when the coefficient
of the term $\partial_i\varphi\partial_i\varphi$ vanishes, $a_1=0$. This is the case of main interest because as we saw the equivalence
with the competing spherical model is established at the Lifshitz point. The results can be related to the critical behavior of
Section \ref{four}.

By adopting the Pauli-Villars regularization, the integrand of (\ref{5.9a}) is replaced by  its regularized expression
involving the regulator $\Lambda$:
\begin{equation}
\frac{1}{2g(\Lambda)}=\int \frac{dk_{0}}{2\pi}\frac{d^{d}k}{(2\pi)^{d}}\left[\frac{i}{k_{0}^{2}-a_{2}^2({\bf k}^{2})^2-m^{4}}-
\frac{i}{k_{0}^{2}-a_{2}^2({\bf k}^{2})^2-\Lambda^{4}}\right].
\label{5.10}
\end{equation}
Isolating the divergent part, by using the dimensional regularization as an intermediate step, we obtain
\begin{equation}
\frac{1}{2g(\Lambda)}-\frac{h(d)}{a_2^{d/2}}\Gamma\left(\frac{2-d}{4}\right)(m^{d-2}-\Lambda^{d-2})+\,\text{finite}=0,
\label{5.11a}
\end{equation}
where $h(d)\equiv\frac{\Gamma(d/4)}{2^{d+2}\pi^{(d+1)/2}\Gamma(d/2)}$ is a positive function
and \textquotedblleft finite\textquotedblright denotes terms which are finite as $\Lambda\to \infty$. To absorb the divergent part,
we introduce the renormalized coupling constant $g_{R}$, defined at some positive mass scale $\mu$, according to
\begin{equation}
\frac{1}{2g(\Lambda)}=\frac{1}{2g_R}+\frac{h(d)}{a_2^{d/2}}\Gamma\left(\frac{2-d}{4}\right)(\mu^{d-2}-\Lambda^{d-2})-\,\text{finite}=0,
\label{5.12a}
\end{equation}
so that the relation between the renormalized coupling constant $g_R$ and the mass $m$ reads
\begin{equation}
\frac{1}{2g_R}-\frac{h(d)}{a_2^{d/2}}\Gamma\left(\frac{2-d}{4}\right)(m^{d-2}-\mu^{d-2})=0.
\label{5.13a}
\end{equation}
At this point it is convenient to introduce a dimensionless coupling constant $\lambda_d$ as $\lambda_d\equiv \mu^{d-2}g_R$.
Thus the relation above can be written in terms of $\lambda_d$,
\begin{equation}
\frac{1}{2\lambda_d}-\frac{h(d)}{a_2^{d/2}}\Gamma\left(\frac{2-d}{4}\right)\left[\left(\frac{m}{\mu}\right)^{d-2}-1\right]=0.
\label{5.14a}
\end{equation}
The mass ratio can be isolated according to
\begin{equation}
\left(\frac{m}{\mu}\right)^{d-2}=\frac{1}{\lambda_d}(\lambda_d-\lambda_d^c),
\label{5.15a}
\end{equation}
with the critical coupling constant $\lambda_d^c$ defined as
$\lambda_d^c\equiv \frac{-a_2^{d/2}}{2h(d)\Gamma\left(\frac{2-d}{4}\right)}$. Notice that for the case
$d=2$ the critical coupling constant vanishes. Another observation is that as the parameters $m$ and $\mu$ are positive, we must have $\lambda_d>\lambda_d^c$.
In fact, we are considering here only the case where the $O(N)$ symmetry is not broken, $\lambda_d >\lambda_d^c$.
When $\lambda_d <\lambda_d^c$, the symmetry is broken and at least one of components of $\varphi_a$ acquires a nonzero vacuum expectation value, for example,
$\langle \varphi_1\rangle\neq 0$.

From expression (\ref{5.14a}) we may immediately obtain the renormalization group $\beta$-functions depending on the dimension,
\begin{equation}
\beta_d= \mu\frac{\partial g_{R}}{\partial \mu}= -\frac{8h(d)}{a_2^{d/2}}\Gamma\left(\frac{6-d}{4}\right)(\lambda_d-\lambda_d^c)\lambda_d.
\label{5.16a}
\end{equation}
Some observations are in order. At the dimensions we are considering (where the model is renormalizable), namely $d<6$, we have  $\Gamma\left(\frac{6-d}{4}\right)>0$
what implies that the theory is stable in the ultraviolet (remember that $h(d)$ is a positive function of dimension).
As it happens for the relativistic situation (corresponding to $d=z=1$), the case $d=2$ shows that the theory is asymptotically free
and has only the trivial fixed point at origin due to the vanishing of the critical coupling constant. Namely, the $\beta$-function (\ref{5.16a})
reduces to
\begin{equation}
\beta_2= - \frac{1}{2\pi a_2}\,g^{2}_{R}.
\label{5.17a}
\end{equation}
The perturbative calculation of the $\beta$-function in the case $d=z=2$ was
performed in \cite{Farakos}. The result obtained there coincides with the above one after taking the
large-$N$ limit and with an appropriate identification between the parameters of the models.

For the case $d>2$, the $\beta$-functions (\ref{5.16a}) exhibit Wilson-Fisher nontrivial fixed points $\lambda_d^c$,
given by
\begin{equation}
\lambda^c_3=\frac{2a_2^{3/2}\pi^{5/2}}{\left(\Gamma(3/4)\right)^2},~~~\lambda^c_4=16\pi^2 a_2^2,~~~\text{and}~~~
\lambda^c_5=\frac{36\pi^{7/2}a_2^{5/2}}{\Gamma(1/4)\Gamma(5/4)}.
\label{5.18a}
\end{equation}
The existence of nontrivial fixed points is associated with phase transitions. In fact, by
comparing with the equation (\ref{4.2}) of Section \ref{four}, we observe that in $d=2$, where the competing quantum spherical model
does not exhibit a phase transition, the $\beta$-function has only a trivial fixed point.
For $d=3,4$, and 5 the spherical model exhibits a nontrivial critical behavior, which corresponds in the sigma model the
existence of nontrivial fixed points of the $\beta$-functions.
For $d\geq 6$ the spherical model exhibits a mean-field behavior that corresponds to the nonrenormalizability of the nonlinear sigma model.

Before closing this section it is opportune to comment about the relativistic sigma model. It is known that the
model is renormalizable in the $1/N$ expansion in $d=1$ and 2 (remember that $d$ denotes only spatial dimensions). In the case $d=1$, the $\beta$-function
has only a trivial fixed point whereas in $d=2$ it exhibits a nontrivial fixed point \cite{Arefeva1}.
These results can be compared with equation (\ref{4.1}), i.e., outside the Lifshitz point, where $J({\bf q })$ is dominated
by quadratic terms giving rise to an essentially relativistic behavior. We see that the spherical model
does not exhibit phase transition in $d=1$ and has a nontrivial phase transition for $d=2$. For $d\geq 3$, where
the behavior of the spherical model is of mean-field type, it corresponds to the nonrenormalizability of the nonlinear sigma model.

\subsection{The renormalization procedure}

The renormalization scheme that we will employ rests heavily on the graphical identity  depicted  in Fig. \ref{qw1},  first found in the relativistic situation in \cite{Arefeva1,Arefeva2}; it  is a consequence of the
unrenormalized sigma propagator being minus the inverse of the amplitude associated with the bubble diagram.
\begin{figure}[!h]
\centering
\includegraphics[scale=0.7]{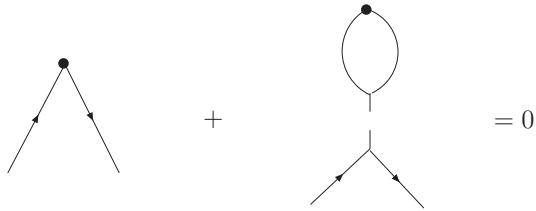}
\caption{A basic identity. It is a consequence from the fact that     the $\sigma$ field propagator is minus the inverse of the bubble diagram.}
\label{qw1}
\end{figure}
It is important to notice that the graphical identity remains valid even  if there is in the integrand  a  factor linear in the momentum carried by one of the lines in the loop. This is so because
\begin{equation}
\int \frac{dk_{0}}{2\pi}\frac{d^{d}k}{(2\pi)^{{d}}} \, k_{\mu}\Delta(k) \Delta(-k+ p)=\frac{ p_{\mu}}2 \int \frac{dk_{0}}{2\pi}\frac{d^{d}k}{(2\pi)^{{d}}}\,\Delta(k) \Delta(-k+ p)=-\frac{ p_{\mu}}{2\Delta_{\sigma}(p)}\label{5.21},
\end{equation}
where $\Delta_{\sigma}$ is the sigma field propagator and for simplicity we omitted the $O(N)$ indices.
This result may be used to prove the cancellation of divergences on some graphs. Take for example the $1/N-$leading contribution to the four point function $\langle  {\rm T}\varphi_{a}\varphi_{a}\varphi_{b}\varphi_{b}\rangle$ where $a$ and $b$ are different $O(N)$ indices; its graphical representation is  shown in Fig. \ref{Ex}d.  It is then found that, once all  the graphs in Fig. \ref{Ex} are taken into account,  there is a complete cancellation of subtraction  terms generated  by the application of the second order Taylor operator
around zero external momenta (see details in the Appendix). For a discussion of subtraction of divergences with
Taylor operators and BPHZ renormalization procedure in anisotropic theories see \cite{Gomes,Gomes1}.

We will now examine the renormalization parts (proper graphs with nonnegative degree of divergence) contained in a arbitrary diagram. There are the following possibilities:

1. Graphs with $N_{\sigma}=1$ and $N_{\varphi}\geq 2$. For $d=2$ irrespectively of $N_{\varphi}$, any graph will be logarithmically divergent. For $d>2$ only diagrams with $N_{\varphi}=2$ will be (logarithmically) divergent;

2. Graphs with $N_{\sigma}=0$ and $N_{\varphi}\geq 6$. Regardless of $N_{\varphi}$,  for $d=2$ any graph will be quartically divergent.  For $d>2$ the divergence of an arbitrary graph may be  at most quadratic;

3. Graphs with $N_{\sigma}=0$ and $N_{\varphi}=4$. Taking aside the case $d=2$ where the divergences are quartic,  the worst divergence is cubic and occurs for $d=3$; nevertheless, due to the rotational symmetry, subtraction terms containing odd powers of the external momenta in amplitudes adequately regularized do not require counterterms. Thus, if regularized the global divergences appear in the coefficient
of a polynomial in the external momenta of at most second order degree;

4. Graphs with $N_{\sigma}=0$ and $N_{\varphi}=2$. Here for any dimension that we have been considering every graph is  quartically divergent. The overall subtraction  may be done by applying the Taylor operator $ t^{2,4}_{p_0,\bf p }$
of second order in $ p_{0} $ and fourth order in $ \bf p $, where $p_{0}$ and $\bf p$ are the components of the external momentum. Aside a possible mass counterterm,  these subtractions generate
counterterms proportional to $ \partial_{0}\varphi \,\partial_{0}\varphi$, $ \partial_{i}\varphi \,\partial_{i}\varphi$
and $ \Delta \varphi\Delta \varphi $; they may be generated by a wave function renormalization and reparameterizations of the
couplings  of the  terms with two and four space derivatives. As we will see shortly, the mass counterterm is not necessary as it can be removed  by adjusting  the parameters in the renormalized Lagrangian.

With the exception of the cases 1. and 4., where reparametrizations of the original Lagrangian automatically furnish the needed counterterms, the cases
2. and 3. need  special consideration. We now argue that no additional counterterm is necessary if we restrict ourselves to the Green function of the
$\varphi$ field (no external sigma lines).

To verify this result, let us consider  a generic diagram $G $ as the one  in Fig. \ref{qw2}, where the hatched bubble represents a graph irreducible with respect to all fields, without external sigma lines and having $N_{\varphi} \geq 6$ .
\begin{figure}[!h]
\centering
\includegraphics[scale=0.6]{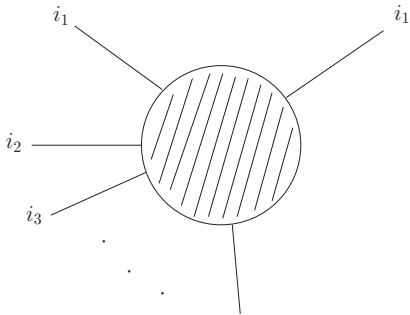}
\caption{A generic diagram. The hatched bubble represents a graph proper with respect to all fields.}
\label{qw2}
\end{figure}
We suppose that all proper subgraphs of $G$ have been made finite by subtracting their divergences according the BPHZ forest formula
and will prove now that all these subtractions
cancels. Indeed, as the maximum divergence is quartic and there are at least $6$ external lines, it is always possible to find  in the subtraction terms a pair of lines
with the same $O(N)$ index and carrying  momenta which appears at most linearly in the subtraction operator.
To $G$ we associate  an expanded diagram $\bar G$ obtained from $G$ by joining two external lines carrying the same $O(N)$ index in a $\sigma \varphi^{2}$ vertex and by attaching  two external lines to a new $\sigma \varphi^{2}$ vertex linked to the first vertex by a $\sigma$ line (see
Figs. \ref{qw2} and \ref{qw3}). This expanded diagram  has the same diagram $G$ as its largest divergent subgraph and by construction $G$ and $\bar G$ have the same order in $1/N$.
Notice now that, because of  Fig. \ref{qw1}, the reduced diagram $\bar G/G$ (recall that a reduced diagram $G/\gamma$ is the graph obtained by contracting the subgraph $\gamma$ of $G$ to a point) is precisely the one associated with the corresponding subtraction for $G$. The amplitude for the reduced diagram has however an additional minus sign and so cancels with the subtraction for the graph $G$.

\begin{figure}[!h]
\centering
\includegraphics[scale=0.6]{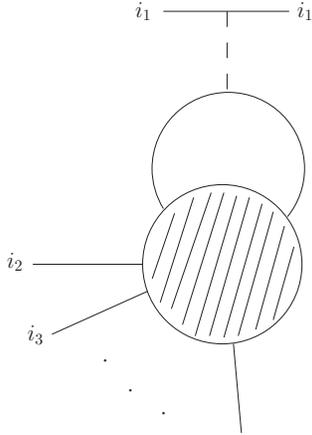}
\caption{An expanded diagram associated with the graph \ref{qw2}.}
\label{qw3}
\end{figure}

To conclude  our analysis, let us now look at the situation 3. in the list above, the four point function for the $\varphi$ field. If $d\not= 2$ it is also possible to find a pair of lines with the same $O(N) $ indices  and carrying momenta which appear at most linearly in the subtraction terms. Thus the same construction as in the case before applies and no counterterm is necessary.

 Differently from the previous case, if $d=2$  the graphs are quartically divergent and it is possible  to have subtraction terms in which each pair of lines with the same $O(N)$ index carries more than one momentum factor coming from the subtraction operator.  They will require counterterms with four derivatives,
as $(\varphi_{a} \triangle \varphi_{a})(\varphi_{b} \triangle \varphi_{b})$ and  $(\varphi_{a} \partial_{i} \partial_{j}\varphi_{a})(\varphi_{b}\partial_{i} \partial_{j}  \varphi_{b})$.

To sum up,  in accord with the above scheme most of the divergences are automatically canceled whereas,  for $d=3,4$ and $5$, the remaining ones
are eliminated   by defining renormalized quantities through the replacements
\begin{eqnarray}
\varphi &\quad\rightarrow\quad & Z^{1/2}_{\varphi}\varphi=(1+a)^{1/2}\varphi\nonumber\\
\sigma &\quad\rightarrow \quad & Z_{\sigma} \sigma=(1+b)\sigma\nonumber\\
a_1 &\quad\rightarrow \quad & Z^{1/2}_{a_{1}} a_{1}= (1+c)^{1/2} a_{1}\nonumber\\
a_{2}&\quad\rightarrow \quad & Z^{1/2}_{a_{2}} a_{2}=(1+ d)^{1/2} a_{2}\nonumber\\
1/g &\quad\rightarrow \quad & Z_{g}/g =(1+f)/g.
\end{eqnarray}
For simplicity of notation we are considering the same letters for the renormalized parameters.
The Lagrangian takes the form,
\begin{equation}
{\cal L}= {\cal L}_{un}+ {\cal L}_{ct},
\end{equation}
with the counterterm Lagrangian  given by
\begin{equation}
{\cal L}_{ct}= \frac{a}2 \partial_{0}\varphi \partial_{0}\varphi -\frac{B}2 \partial_{i}\varphi \partial_{i}\varphi +\frac{C}2 \partial_{i}\partial_{j}\varphi \partial_{i}\partial_{j}\varphi-\frac{m^{4} a}2\varphi^{2} -\frac{ D}{\sqrt{2N}} \sigma \varphi^{2}+ F\sqrt{\frac{N}2}\frac{\sigma}{2g},
\end{equation}
where  we introduced
\begin{eqnarray}
B&=& (1+a)(1+c) -1\nonumber\\
C&=& (1+a)(1+d)-1\nonumber\\
D&=&(1+a)(1+b)-1\nonumber\\
F&=& (1+b)(1+f)-1.
\end{eqnarray}

As remarked before and can be straightforwardly verified,  the mass counterterm is innocuous  since, due to the identity of Fig. \ref{qw1},  it cancels in the contributions to the Green functions. The surviving divergences  in the two point function of the $\varphi$ field are then eliminated by adjusting the counterterms with derivatives, those with coefficients $a$, $B$ and $C$ in ${\cal L}_{ct}$. Notice also that the $F$ counterterm may be chosen as to eliminate higher orders tadpoles and ensure that $m$  is the \textquotedblleft physical\textquotedblright   mass in the sense that
\begin{equation}
\Gamma^{(2)}(k)= 0, ~~  \text{for}~~ k_{0}^{2}= a_{1}{\bf k}^{2}+a_{2}({\bf k}^{2})^2+m^{4},
\end{equation}
where $\Gamma^{(2)}(k)$ is the two point 1PI (with respect to the $\varphi$ fields) vertex function.
Similarly, the $D$ counterterm is also irrelevant as far the Green functions of the $\varphi$ field are concerned. Actually, since
only the $\varphi$ field has a physical interpretation for renormalization purpose any graph containing external lines of the sigma field will be considered  just as a subgraph of large graphs without external sigma lines.

An special situation arises at $d=2$ which, for consistency, requires  the introduction of new interaction terms in the Lagrangian  (\ref{5.9}) so that by reparametrizations the needed counterterms are produced. These are $O(N) $ invariant composite operators made of just four basic fields and their derivatives of the form
\begin{equation}
\lambda_{1}(\varphi_{a} \triangle \varphi_{a})(\varphi_{b} \triangle \varphi_{b})+\lambda_{2}(\varphi_{a} \partial_{i} \partial_{j}\varphi_{a})(\varphi_{b}\partial_{i} \partial_{j}  \varphi_{b}).
\end{equation}





\section{Summary}\label{Summary}

We investigated various statistical mechanical and field theoretical aspects arising from
the connection between the continuum limit of quantum spherical model with
competing interactions and the Lifshitz-type $O(N)$ nonlinear sigma model with $N$ tending to infinity.

We started by discussing some features of the quantum spherical model with
competing interactions. Certain relations between the parameters $J_1$, $J_2$, and $J_3$ were determined
by considering issues as Lifshitz point and rotational symmetry.
The maximum of $J({\bf q})$ depends on a special combination of the interaction energies defined through
the parameter $p\equiv \frac{-[ J_2+\frac12(d-1)J_3]}{J_1}$.
For $p\leq 1/4$  its maximum is given by ${\bf q}_c=0$, whereas for $p> 1/4$ we have ${\bf q}_c\neq 0$. The
point $p=1/4$ separates these two regions characterizing the Lifshitz point. At the Lifshitz point the system exhibits
an anisotropic behavior between space and time coordinates.

The rotational symmetry reduces the number of independent
parameters since it relates the antiferromagnetic couplings $J_2$ and $J_3$, according to $J_3=2J_2$.
This situation is important mainly when we take the continuum limit in order to identify the underlying rotational invariant field theory with
the presence of higher spatial derivative terms.

In the quantum derivation of the equivalence between the models, our strategy was to analyze the large-$N$ quantum effective action of the
anisotropic nonlinear sigma model and then to take the limit $N\rightarrow\infty$. We ended up with the gap equation,
that when taking into account the temperature reduces to the saddle point condition of the quantum spherical model.

Regarding the critical behavior of the quantum spherical model at zero temperature, we found the critical dimensions of the model
by analyzing the saddle point condition nearby the critical point.
Specifically, we determined the lower critical dimension, where the quantum fluctuations are
too strong preventing the formation of an ordered state, and
the upper critical dimension, above which the fluctuations are not relevant anymore and the system has a typical
mean-field behavior. Between these two dimensions the system exhibits a nontrivial critical behavior.
At the Lifshitz point, $p=1/4$, the  lower and upper critical dimensions are $d_l=2$ and $d_u=6$, respectively. In this situation we
have a nontrivial critical behavior in dimensions $d=3,4$, and $5$. These results were compared with the fixed point
structure of the $\beta$-functions of the anisotropic nonlinear sigma model in the corresponding dimensions.
Outside the Lifshitz point, $p\neq 1/4$, the critical dimensions are $d_l=1$ and $d_u=3$, and we have a nontrivial
critical behavior in $d=2$, what corresponds to the nontrivial fixed point in the relativistic nonlinear sigma model in $2+1$ spacetime dimensions.

We formulated general anisotropic nonlinear sigma model for arbitrary values of $z$ and the conditions for large-$N$ renormalizability
by systematic power counting depending on $z$ and $d$. We then restricted our attention to the case $z=2$,
studying in detail the renormalization procedure in the context of the $1/N$ expansion and performing the calculation of the
$\beta$-functions in $d=2,3,4$, and $5$, the dimensions in which the model is renormalizable.
In the case $d=2$, there is only the trivial fixed point which corresponds to the non existence of phase transitions
in the quantum spherical model, i.e., when we are considering the system at the lower critical dimension.
In the cases, $d=3,4$, and $5$, on the other hand, we found Wilson-Fisher nontrivial fixed points
corresponding to the existence of nontrivial critical behavior in the
quantum spherical model at the Lifshitz point.

Concerning the renormalization of the model, we analyzed the structure of
divergent 1PI Green functions. The majority of the divergences are automatically canceled in the $1/N$ expansion while
the remaining ones are absorbed in the redefinition
of the parameters of the theory. Only for the case $d=2$ was necessary to add to the lagrangian counterterms proportional
to  $(\varphi_{a} \triangle \varphi_{a})(\varphi_{b} \triangle \varphi_{b})$ and
$(\varphi_{a} \partial_{i} \partial_{j}\varphi_{a})(\varphi_{b}\partial_{i} \partial_{j}  \varphi_{b})$.
These are renormalizable vertices that because of the graphical identity in Fig. \ref{qw1} do not generate additional counterterms.


\section{Acknowledgments}

The authors thank Professor Silvio Salinas and Masayuki Hase for reading the manuscript, useful comments, and suggestions.
This work was partially supported by  Conselho
Nacional de Desenvolvimento Cient\'{\i}fico e Tecnol\'ogico (CNPq) and Funda\c{c}\~ao de Amparo a Pesquisa do Estado de S\~ao Paulo (FAPESP).


\begin{figure}[!h]
\centering
\includegraphics[scale=0.7]{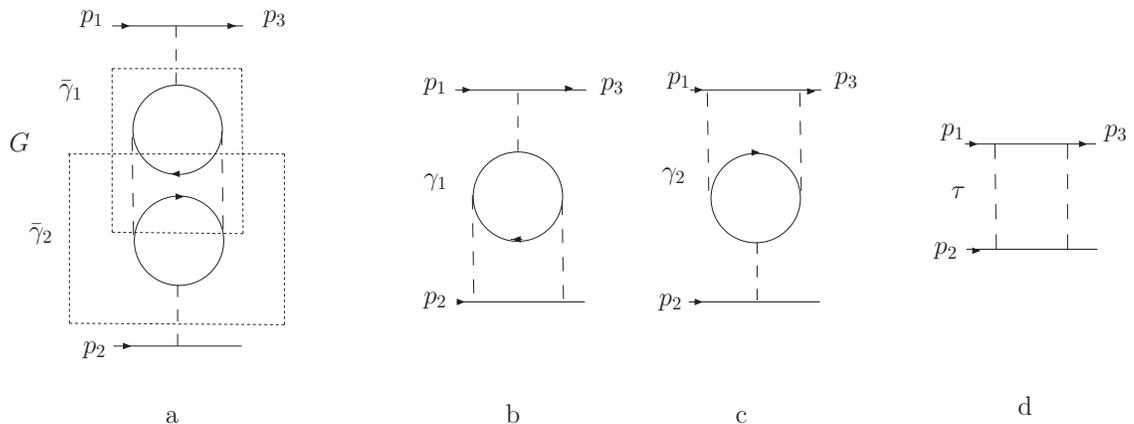}
\caption{Nontrivial leading contributions to the four point function of the $\varphi$ field.}
\label{Ex}
\end{figure}


\appendix

\section{An illustrative example}

In this appendix we will illustrate the general mechanism of cancellation of divergences by analyzing the leading  contributions to the Green function $\langle \text{T}\varphi_{a}\varphi_{a}\varphi_{b}\varphi_{b}\rangle$ with $a\not=b$. The example involves the diagrams $G$, $\gamma_{1}$, $\gamma_{2}$ and $\tau$,
shown in Fig. \ref{Ex}, which are of the same order in $1/N$ although individually they have different number of loops. We will concentrate on space dimensions
$d$ greater than two but less than six. In these situation diagrams $\gamma_{1}$ and $\gamma_{2}$ are both logarithmically divergent whereas $\tau$
has degree of superficial divergences equal to $6-d$.  To cope with the subtraction terms separately,  we suppose that all integrals are dimensionally regularized so that the actual physical dimension $d$ is taken at the end of the calculation.  Diagram $G$ has $\bar \gamma_{1}$, $\bar\gamma_{2}$ and $\tau$ as subgraphs
and therefore in the BPHZ scheme presents the following forests $\emptyset$, $\bar \gamma_{1}$, $\bar\gamma_{2}$, $\tau$, $\{\bar\gamma_{1},\,\tau\}$ and $\{\bar\gamma_{2},\,\tau\}$. Thus the subtracted integrand for the diagram $G$ is given by
 \begin{equation}
R_G= I_G- I_{G/
\bar\gamma_1}t^{0}_{\bar\gamma_1} I_{\bar\gamma_1}- I_{G/\bar\gamma_2}
t^{0}_{\bar\gamma_2} I_{\bar\gamma_2}- I_{G/\tau}t^{6-d}_{\tau} I_{\tau}+
I_{G/\bar\gamma_1}t^{0}_{\bar\gamma_1}I_{\bar\gamma_1/\tau}t^{6-d}_{\tau} I_{\tau}+
I_{G/\bar\gamma_2}t^{0}_{\bar\gamma_2}I_{\bar\gamma_2/\tau}t^{6-d}_{\tau} I_{\tau},
\end{equation}
where $I_G$ denotes the unsubtracted amplitude associated with the graph $G$. Similarly, the amplitudes associated with the graphs $\gamma_{1}$,
$\gamma_{2}$ and $\tau$ are, respectively,
\begin{eqnarray}
R_{\gamma_1}&=& I_{\gamma_1}-
I_{\gamma_1/\bar\gamma_1}t^{0}_{\bar\gamma_1} I_{\bar\gamma_1} - I_{\gamma_1/\tau}
t^{6-d}_{\tau} I_\tau+ I_{\gamma_1/\bar\gamma_1}t^{0}_{\bar\gamma_1}
I_{\bar\gamma_1/\tau}t^{6-d}_{\tau} I_{\tau},\\
R_{\gamma_2}&=& I_{\gamma_2}-I_{\gamma_2/\bar\gamma_2}
t^{0}_{\bar\gamma_2} I_{\bar\gamma_2} - I_{\gamma_2/\tau}
t^{6-d}_{\tau} I_\tau+ I_{\gamma_2/\bar\gamma_2} t^{0}_{\bar\gamma_2}
I_{\bar\gamma_2/\tau}t^{6-d}_{\tau} I_{\tau},
\end{eqnarray}
and
\begin{equation}
R_\tau= I_\tau- t^{6-d}_{\tau} I_\tau.
\end{equation}

By the use of the identity in Fig. \ref{qw1}, we may immediately  cancel (after integrating on loop momenta)  various terms in the sum of the subtracted amplitudes such that, effectively,
\begin{eqnarray}
R_{G}+R_{\gamma_{1}}+R_{\gamma_{2}} +R_{\tau}&=& I_{G}+I_{\gamma_{1}}+I_{\gamma_{2}} +I_{\tau}- I_{G/\tau}t_{\tau}^{6-d}I_{\tau}\nonumber\\
&&-I_{\gamma_{1}/\tau}t_{\tau}^{6-d}I_{\tau}-I_{\gamma_{2}/\tau}t_{\tau}^{6-d}I_{\tau}-t^{6-d}_{\tau} I_\tau.
\end{eqnarray}
To describe the action of the derivatives in $t_{\tau}^{6-d}$ we will represent by a small circle on a line the effect of one derivative applied to
the propagator associated with the line.
Using this representation, notice that the subtraction terms in which the derivatives in $t_{\tau}^{6-d}$ act just on the upper line or just in the lower line of
$\tau$ cancel among themselves. In fact, suppose that two derivatives with respect to the external momenta act on the upper line of Fig. \ref{Ex}d, as  indicated in Fig. \ref{Scheme2}a. Then, it is easy to see that this contribution will be cancelled by the one coming from the  Fig. \ref{Scheme2}b.

It remains to analyze the cases in which there are derivatives acting both in the upper and lower lines of $\tau$. By symmetric integration, we need to consider only the situation where there are  two derivatives, one with respect to the momentum in the upper line and the other with respect to the momentum in the lower line.  The cancellation here is more complicated due the occurrence of a momentum factor in the integrand of reduced
graphs which produces an additional factor of $1/2$, as indicated in  equation (\ref{5.21}). In Fig. \ref{Scheme} we have redrawn the diagrams of Fig. \ref{Ex} displaying
a specific choice for the  route of  the external momenta (but omitting the loop momenta); we  will verify the cancellation of the subtraction terms proportional to $p_{1}p_{2}$ or
to $p_{2}p_{3}$.  The sum of  the contributions coming from graphs \ref{Scheme}a and \ref{Scheme1}a give the result which, because of the  identity in equation (\ref{5.21}),
 vanishes when the sum of the contributions coming from the \ref{Scheme}b and \ref{Scheme1}b is taken into consideration. The same happens with the sum of contributions coming from \ref{Scheme}c and \ref{Scheme1}c which is cancelled by the sum of \ref{Scheme}d with \ref{Scheme1}d.

\begin{figure}[!h]
\centering
\includegraphics[scale=0.7]{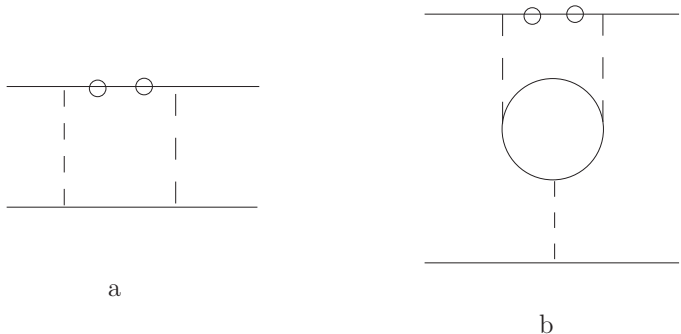}
\caption{Example of the cancellation of subtraction terms when the derivatives are just with respect the momenta in the upper line. The small circle denotes a derivative.}
\label{Scheme2}
\end{figure}

\begin{figure}[!h]
\centering
\includegraphics[scale=0.7]{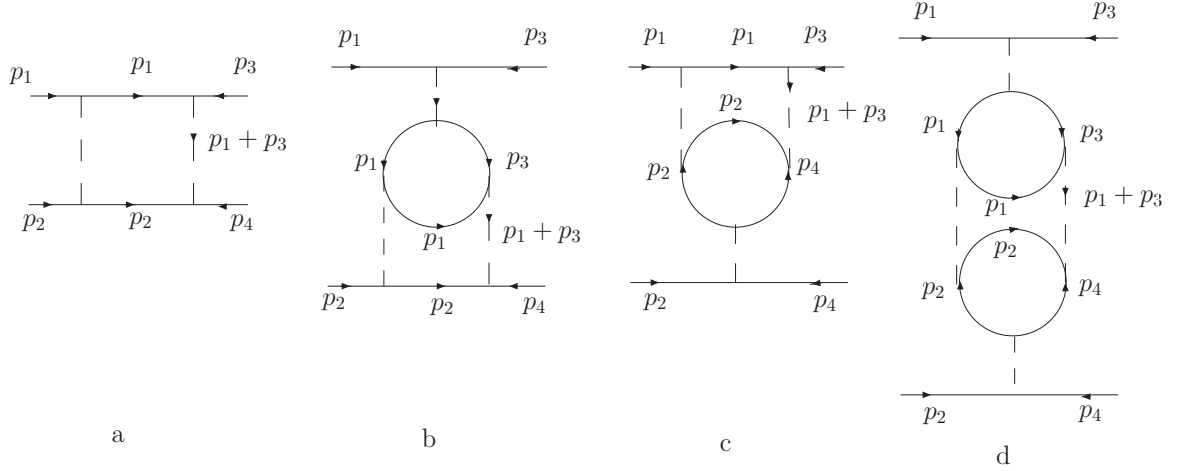}
\caption{Flow of external momenta in the "direct"  channel ($p_{1}$ and $p_{2}$ entering at vertices linked by just one sigma line. }
\label{Scheme}
\end{figure}

\begin{figure}[!h]
\centering
\includegraphics[scale=0.7]{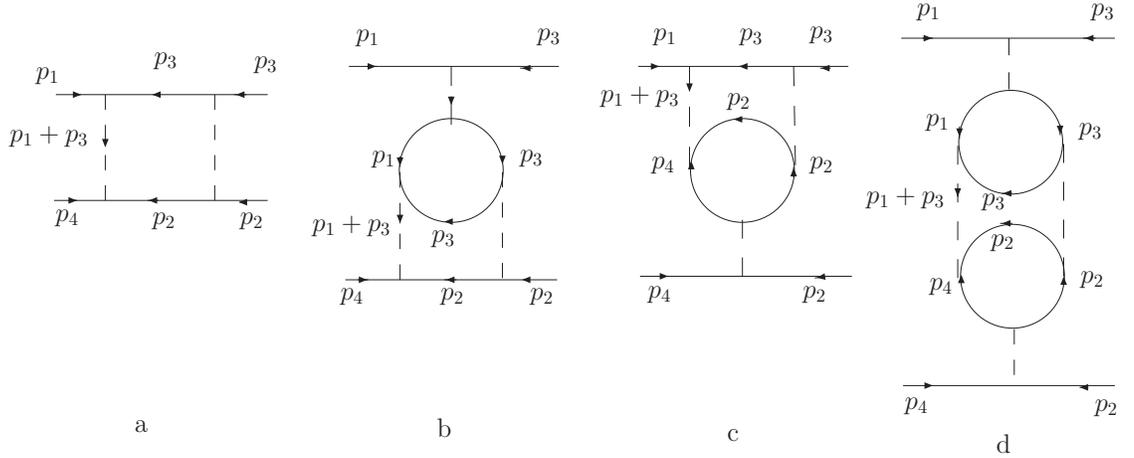}
\caption{Flow of the external momenta in the "crossed" channel (same as in Fig. \ref{Scheme} but with $p_{2}$ and $p_{4}$ exchanged.}
\label{Scheme1}
\end{figure}


\end{document}